\documentclass[journal]{IEEEtran}
\usepackage{cite}

\ifCLASSINFOpdf
\usepackage{romannum}
   \usepackage[pdftex]{graphicx}
   \usepackage{caption}
   \usepackage{dcolumn}
\else
\fi
\usepackage{amsmath}
\usepackage{bm}
\usepackage{amssymb}
\usepackage{stmaryrd}
\usepackage{float}
\newcommand{\dy}[1]{\underline{\underline{\bm{#1}}}}

\newcommand*{\rom}[1]{\expandafter\@slowromancap\romannumeral #1@}
 \usepackage[caption=false,font=normalsize,labelfont=sf,textfont=sf]{subfig}

\usepackage{stfloats}
\begin{document}
%
\title{Spectral-Domain Method of Moments Analysis of Spatially Dispersive Graphene Patch Embedded in Planarly Layered Media}
%
%
%

\author{Minyu Gu,~\IEEEmembership{Student Member,~IEEE,}
	and Krzysztof A. Michalski,~\IEEEmembership{Life~Fellow,~IEEE}
	\thanks{Minyu Gu (e-mail: guminyu@tamu.edu) and Krzysztof A. Michalski (e-mail: k-michalski@tamu.edu) are with the Department of Electrical and Computer Engineering, Texas A\&M University, College Station, TX, 77843 USA.}
	\thanks{Manuscript received March 28, 2022; revised March 28, 2022.}}

%
%

\markboth{AP2203-0567}%
{Shell \MakeLowercase{\textit{et al.}}: Spectral-Domain Method of Moments Analysis of Spatially Dispersive Graphene Patch Embedded in Planarly Layered Media}
%



\maketitle

\begin{abstract}
 \textit{ Anisotropic and spatially dispersive graphene patches of arbitrary shape embedded in planarly layered uniaxial media are analyzed using spectral-domain method of moments. Formulation and computational methods for the spectral-domain method of moments using the Rao-Wilton-Glisson subdomain basis function and incorporating the full-wavevector Bhatnagar-Gross-Krook formulation of graphene surface conductivity tensor are proposed.} The impedance matrix is efficiently evaluated by a novel numerical method which firstly approximates the spectral-domain Green function, basis function and conductivity tensor with Chebyshev polynomials, and then sums up the Fourier transformed coefficients. \textit{Blue-shift of the resonant frequency and variation of the current distribution due to spatial dispersion are observed in various structures demonstrated.}
\end{abstract}

%
\IEEEpeerreviewmaketitle
\section{Introduction}
\IEEEPARstart{S}{urface} plasmonic waves (SPWs) \textit{that} propagate along graphene and other two-dimensional (2D) materials have recently attracted significant research interests. It has been shown that the intrinsic plasmons frequency of graphene lies in the low TerraHerz (THz) frequency regime, in contrast with noble metals only allowing plasmons to exist in the visible regime, which promises a significant breakthrough for unconventional plasmonic devices\cite{ju2011graphene}. Due to the \textit{atomically}  thin nature of the 2D materials, for macroscopic simulation, it is common to use a surface conductivity tensor to model the electromagnetic waves and plasmons coupling in theoretical studies. However, graphene is known to be spatially dispersive\cite{falkovsky2007space}, which is related to the spatial variation of the charge carrier number density and implies the \textit{conductivity} tensor to vary with wavevectors. This important characteristic \textit{has usually been} neglected in works related to SWPs device\cite{hanson2008dyadic}, but recently proved to be essential in many scenarios. For instance, \textit{it is reported} in \cite{lovat2015nonlocal} and \cite{silveiro2015quantum} that the spatial dispersion (also known as nonlocal) effects of graphene striplines will result in the resonant frequency and current distribution departing from the local or low-wavevector approximation model if the \textit{ribbon} width is narrow. Moreover, the introduction of active SPWs reveals that nonreciprocity of electromagnetic waves can be induced via applying DC bias current on the graphene\cite{morgado2020nonlocal}. Interestingly, if the phase velocity of SPWs exceeds the bias electrons velocity, active \textit{travelling-wave} amplification of SPWs can be achieved on the graphene plasmonic platform\cite{ghafarian2016millimetre}. Therefore, accurate modeling of the spatial dispersion effects is vital for the simulation of both passive and active 2D nano-device.
\par
The full-wave modeling of graphene sheets and ribbons partially incorporating spatial dispersion has been previously studied using the dyadic Green function formulation \cite{hanson2008dyadic} and 2D method of moments (MOM) with entire-domain basis functions \cite{lovat2015nonlocal}. Also, a discontinuous Galerkin time-domain framework \cite{li2018discontinuous} and a spatial-domain 2D-MOM with second-order approximation \cite{burghignoli2014space} \textit{have been} developed to incorporate the low-wavevector approximation of the graphene conductivity. \textit{In the present article} we propose a spectral-domain MOM with subdomain triangular Rao-Wilton-Glisson (RWG) basis functions \cite{rao1982electromagnetic} that incorporates the full-wavevector Bhatnagar-Gross-Krook formulation of graphene conductivity \cite{lovat2013semiclassical}, which is refered as nonlocal model in this articale in contract with the local model which evaluates the conductivity as zero-wavevector value. To facilitate the computation of the impedance matrix, the conductivity tensor and spectral-domain dyadic Green function are evaluated by computing a 2D inverse Fourier integral. This integral is further accelerated by firstly approximating the functions in the integrand with Chebyshev polynomials, and then \textit{summing up} the Fourier transformed Chebyshev coefficients. This procedure not only achieves substantial speed-ups compared with a direct numerical \textit{integration}, but is also insensitive to the spatial distance of the basis and testing functions.
\par
\textit{To demonstrate the capability of the proposed method to model planar graphene patches of arbitrary shape embedded in planarly layered uniaxial media, several structures, graphene dipole antenna, spiral antenna, two-way power divider, and circular patches are computed. The necessity of incorporating the spatial dispersion effect is preonunced in all the cases except the first one, where the current distribution, port input impedance, and resonact frequency are evidently different between local and nonlocal conductivity models.} 
\par
\section{Theory and Methodology}
\subsection{Problem Definition and Integral Equations Formulation}
We consider an infinitely thin graphene patch of arbitrary shape located on $z$ in layered uniaxial planar layered media. The graphene sheet may be in-plane anisotropic or spatially dispersive, characterized by a surface conductivity tensor
\begin{equation}
\dy{\bm{\sigma}}(k_x,k_y)
=
\begin{bmatrix}
\sigma_{xx} & \sigma_{xy}\\
\sigma_{yx} & \sigma_{yy}
\end{bmatrix}\,.
\label{sigxy}
\end{equation}
The full-wavevector formulation of graphene conductivity is described in Appendix~A. The layered media are assumed to be of infinite lateral extent on transverse xy-plane, invariant in any plane transverse to the z-axis, and may be uniaxially anisotropic. The relative permittivity and permeability dyadic within the n-th layer are characterized by
\begin{align}
\underline{\underline{\bm{\varepsilon}_n}}= \epsilon_{tn} ( \boldsymbol{\hat{x}}\boldsymbol{\hat{x}} +  \boldsymbol{\hat{y}}\boldsymbol{\hat{y}} ) + \epsilon_{zn} \boldsymbol{\hat{z}}\boldsymbol{\hat{z}}\\
\underline{\underline{\bm{\mu}_n}}= 
\mu_{tn} ( \boldsymbol{\hat{x}}\boldsymbol{\hat{x}} +  \boldsymbol{\hat{y}}\boldsymbol{\hat{y}} ) + 
\mu_{zn} \boldsymbol{\hat{z}}\boldsymbol{\hat{z}}\,
\label{eq1}\,.
\end{align}
To express the fields in the spectral domain, the Fourier transform of all transverse fields is conducted, which is defined as
\begin{equation}
\tilde{f}(\bm{k}_\rho)=\mathcal{F}(f)=
\int_{-\infty}^{\infty}\int_{-\infty}^{\infty}
f(\bm{\rho}) \,e^{j\bm{k}_\rho\cdot\bm{\rho}} \,dx dy
\label{e2}
\end{equation}
where $\bm{\rho}=\hat{\bm{x}} x+\hat{\bm{y}} y$ and $\bm{k}_\rho=\hat{\bm{x}} k_x+\hat{\bm{y}} k_y$ is the transverse wavevector.
To facilitate the derivation, the transverse fields and current components are expressed in rotated coordinate and defined as
$\bm{\hat{u}}$, $\bm{\hat{v}}$ and $\bm{\hat{z}}$, where
\begin{equation}
\hat{\bm{u}} = \frac{\bm{k}_\rho}{k_\rho},\quad  \hat{\bm{v}} =\frac{\hat{\bm{z}}\times\bm{k}_\rho}{k_\rho}\,.
\end{equation}
The spectral-domain dyadic Green function of planar layered media can be computed from the theory introduced in \cite{michalski2005electromagnetic}. Since the surface currents flow on the graphene exist on the transverse plane, only the transverse elements of the dyadic Green function $\bm{J}_s$ are included, which can be expressed by voltage fields associated with a transmission line (TL) analogy equations \cite{michalski2005electromagnetic}
\begin{align}
\tilde{\dy{G}}^{EJ}(k_\rho) = 
\begin{bmatrix}
	-V^e_i & 0 \\
	0 & -V^h_i 
\end{bmatrix}
\end{align}
where the $V^e_i, V^h_i$ voltage fields above are solutions of the TL, representing the electric fields generated by the transverse Hertzian dipoles in rotated coordinate. The conductivity tensor can also be transformed as
\begin{subequations}
	\begin{eqnarray}
	&\dy{\tilde{\sigma}}
	=
	\begin{bmatrix}
	\sigma_{uu} & \sigma_{uv}\\
	\sigma_{vu} & \sigma_{vv}
	\end{bmatrix}
	=\dy{M}^T
	\begin{bmatrix}
	\sigma_{xx} & \sigma_{xy}\\
	\sigma_{yx} & \sigma_{yy}
	\end{bmatrix}
	\dy{M} \\
	&\dy{M}=\frac{1}{k_\rho}
	\begin{bmatrix}
	k_x & -k_y\\
	k_y & k_x
	\end{bmatrix}
	\end{eqnarray}	
\end{subequations}
The total electric fields on a conductive sheet are related by the surface currents in the spectral domain as
\begin{equation}
\tilde{\bm{E}_t}(k_x,k_y,z) = \tilde{\dy{\sigma}}^{-1}(k_x,k_y)\cdot \tilde{\bm{J}}_s(k_x,k_y)\,.
\end{equation} 
Upon combining the total fields, we arrive at the electric field integral equations (EFIE) formulation
\begin{equation} 
\begin{split}
&\bm{E}_{inc} + <\dy{G}^{EJ}; \bm{J}_s> =  \mathcal{F}^{-1}[\,\tilde{\dy{\sigma}}^{-1}\cdot \tilde{\bm{J}}_s\,]
\end{split} 
\label{eqint}
\end{equation} 
After applying the convolution theorem, we can write
\begin{equation} 
\mathcal{F}^{-1}\, [(\tilde{\dy{G}}^{EJ} + \tilde{\dy{\sigma}}^{-1}) \cdot \tilde{\bm{J}}_s\, ]   = \bm{E}_{inc}\,.
\end{equation} 
Finally, Galerkin testing procedure is applied by using the spectral-domain RWG basis functions both as weighting and testing functions. The discretized integral equations can be written as
\begin{equation} 
\begin{split}
\sum_{n}\int_S \tilde{\bm{f}}_m \cdot \mathcal{F}^{-1}\, [ (\tilde{\dy{G}}^{EJ} + \tilde{\dy{\sigma}}^{-1}) \cdot \tilde{\bm{f}}_n\, ] \ dS =  \int_S \bm{f}_m \cdot \bm{E}_{inc}\ dS\,.
\end{split}
\end{equation} 
To obtain the spectral-domain representation of the equations, we use the generalized Parseval’s theorem and arrive at
\begin{equation} 
\begin{split}
&\sum_{n} \int\int\tilde{\bm{f}}_m(-k_\rho)\cdot   (\tilde{\dy{G}}^{EJ} + \tilde{\dy{\sigma}}^{-1}) \cdot \tilde{\bm{f}}_n(k_\rho)\, dk_xdk_y \\&=  4\pi^2 \int_S \bm{f}_m \cdot \bm{E}_{inc}^m\ dS\,.
\end{split}
\label{eqdie}
\end{equation} 
The system of linear equations above can be represented in matrix form
\begin{equation}
[\![ Z_{mn} ]\!]\ [\![ I_n ]\!] = [\![ V_m ]\!]\,.
\end{equation}

\subsection{Evaluation of Spectral-domain Integral for Impedance Matrix Filling}
The two-dimensional Fourier integral in Eq.~\ref{eqdie} is difficult to be evaluated due to the fact that the integrand is oscillatory, which is originated from the phase variation associated with distance between weighting and testing functions. In this article we develop a fast approximation method to evaluate the spectral-domain integral. Our method firstly samples the Green functions and basis functions based on the nodes of the zeros of Chebyshev polynomials. During the matrix filling stage, 2D discrete cosine transform is performed to accelerate the computation of Chebyshev coefficients. The integral results are then obtained by summing up the products of the coefficients and precomputed Fourier transform of Chebyshev polynomials evaluated at a specific offset distance. This method circumvents the direct integration, the computational cost is thereby insensitive to the distance between the weighting and testing functions.
\par
We first isolate the phasor due to the relative distance between the testing and weighting functions. The left side of Eq.~\ref{eqdie} can be rewritten as
\begin{eqnarray}
Z_{mn} = \int\int \underbrace{\tilde{\bm{f}}_m^{*}(-k_\rho)\cdot   (\tilde{\dy{G}}^{EJ} + \tilde{\dy{\sigma}}^{-1})}_{\bm{\varTheta}_m} \cdot\underbrace{ \tilde{\bm{f}}_n^{*}(k_\rho)}_{\bm{\varDelta}_n}\notag\\ e^{j\bm{k}_\rho\cdot(\bm{r^\prime_n-r_m})}\, dk_xdk_y
\label{eqzmnd}
\end{eqnarray}
where $r^\prime_n$ and $r_m$ indicate the central coordinate of the RWG basis function, and * indicates that the shifted basis function coordinated at the origin. The Green function $\dy{\tilde{G}}^{EJ}$, inversion of the conductivity tensor $\dy{\tilde{\sigma}}^{-1}$ , and shifted basis function $\tilde{\bm{f}}_m^*,\, \tilde{\bm{f}}_n^*$ are then sampled on the Chebyshev nodes
 \begin{subequations}
 	\begin{eqnarray}
 		 k_{x}^{i} = \cos( \dfrac{\pi i}{N_{x}})k_{x}^m + k_{x}^p,\quad i = 0,\,1...,\,N_{x}-1\\
 	k_{y}^{j} = \cos( \dfrac{\pi j}{N_{y}})k_{y}^m + k_{y}^p,\quad j = 0,\,1...,\,N_{y}-1\\
 		k_{x,y}^{m} = \dfrac{k_{x,y}^b - k_{x,y}^a}{2},\quad	k_{x,y}^{p} = \dfrac{k_{x,y}^b + k_{x,y}^a}{2}
 	\end{eqnarray}
 \end{subequations}
where $N_{x,y}$ indicates the orders of Chebyshev polynomials approximation, and $k_{x,y}^a,\,k_{x,y}^b$ are the truncating bounds of $k_x, k_y$ domains, which are divided into five subdomains to increase the approximation accuracy.
 \begin{subequations}
	\begin{eqnarray}
\Romannum{1}\quad k_x \in [-k_s,k_s],\quad k_y \in [-k_s,k_s]\\
	 \Romannum{2} \quad	k_x \in [-k_s,k_{max}],\quad k_y \in [k_s,k_{max}]\\\Romannum{3}\quad	k_x \in [k_s,k_{max}],\quad k_y \in [-k_{max},k_s]\\\Romannum{4}\quad	k_x \in [-k_{max},k_s],\quad k_y \in [-k_{max},-k_s]\\\Romannum{5}\quad	k_x \in [-k_{max},-k_s],\quad k_y \in [-k_s,k_{max}]\,.
	 \end{eqnarray}
 \end{subequations}
The $\Romannum{1}$ subdomain is evaluated by Chebyshev quadrature along a detour path to circumvent the poles due to guided wave modes of layered media and branch points \cite{michalski2016efficient}. The rest of the subdomains are evaluated by a summation routine described as follows.
\par
Eq.~\ref{eqzmnd} proposates us to separtely sample the testing and weighting functions as two groups
\begin{eqnarray}
&	\bm{\varTheta}_m^{ij} = \tilde{\bm{f}}_m^*(-k_x^i,-k_y^j)
	\label{eqg1}\\
&	\bm{\varDelta}_n^{ij} =  [\tilde{\dy{G}}^{EJ}(k_x^i,k_y^j) + \tilde{\dy{\sigma}}^{-1}(k_x^i,k_y^j)] \cdot \tilde{\bm{f}}_n^*(k_x^i,k_y^j)\,.
	\label{eqg2}
\end{eqnarray}
In the first stage of the program, which is referred as sampling stage, each basis function of Eq.~\ref{eqg1} and Eq.~\ref{eqg2} are sampled and stored into 2D $N_x \times N_y$ arrays. In the next stage referred as the impedance matrix filling stage, m-th or n-th elements are firstly elementwisely multiplied, and 2D discrete cosine transform is then conducted on the resulting arrays to obtain the Chebyshev polynomial coefficents
\begin{equation}
	c_{mn}^{ij} = \dfrac{1}{N_x N_y} DCT2^{*}(\bm{\varTheta}_m^{ij} \circ \bm{\varDelta}_n^{ij})
\end{equation}
where the * indicates the first raw and column of the results obtained from 2D discrete cosine transform are multiplied by $0.5$. $\circ$ indicates element-wisely dot product. The resulting arrays are the coeffiecents of the i-th and j-th order Chebyshev polynomials
\begin{eqnarray}
	\bm{\varTheta}_m \cdot \bm{\varDelta}_n = \sum_i\sum_j c_{mn}^{ij} T_i(k_x)T_j(k_y)\,.
\end{eqnarray}
The Fourier integral in Eq.~\ref{eqzmnd} then can be represented by the Chebyshev polynomials,
\begin{equation}
Z_{mn} = \mathcal{F}^{-1} (\bm{\varTheta}_m \cdot \bm{\varDelta}_n) \approx\sum_i\sum_j c_{mn}^{ij} \tilde{T}_i(x)\tilde{T}_j(y)\,.
\label{eqzmnsum}
\end{equation}
To obtain the Fourier transform of i-th or j-th order Chebyshev polynomial $\tilde{T}_{i,j}$ evaluated at a specific point x or y, a recursive relation is derived by invoking the integration by parts
\begin{figure}[t]
	\includegraphics[width=\columnwidth]{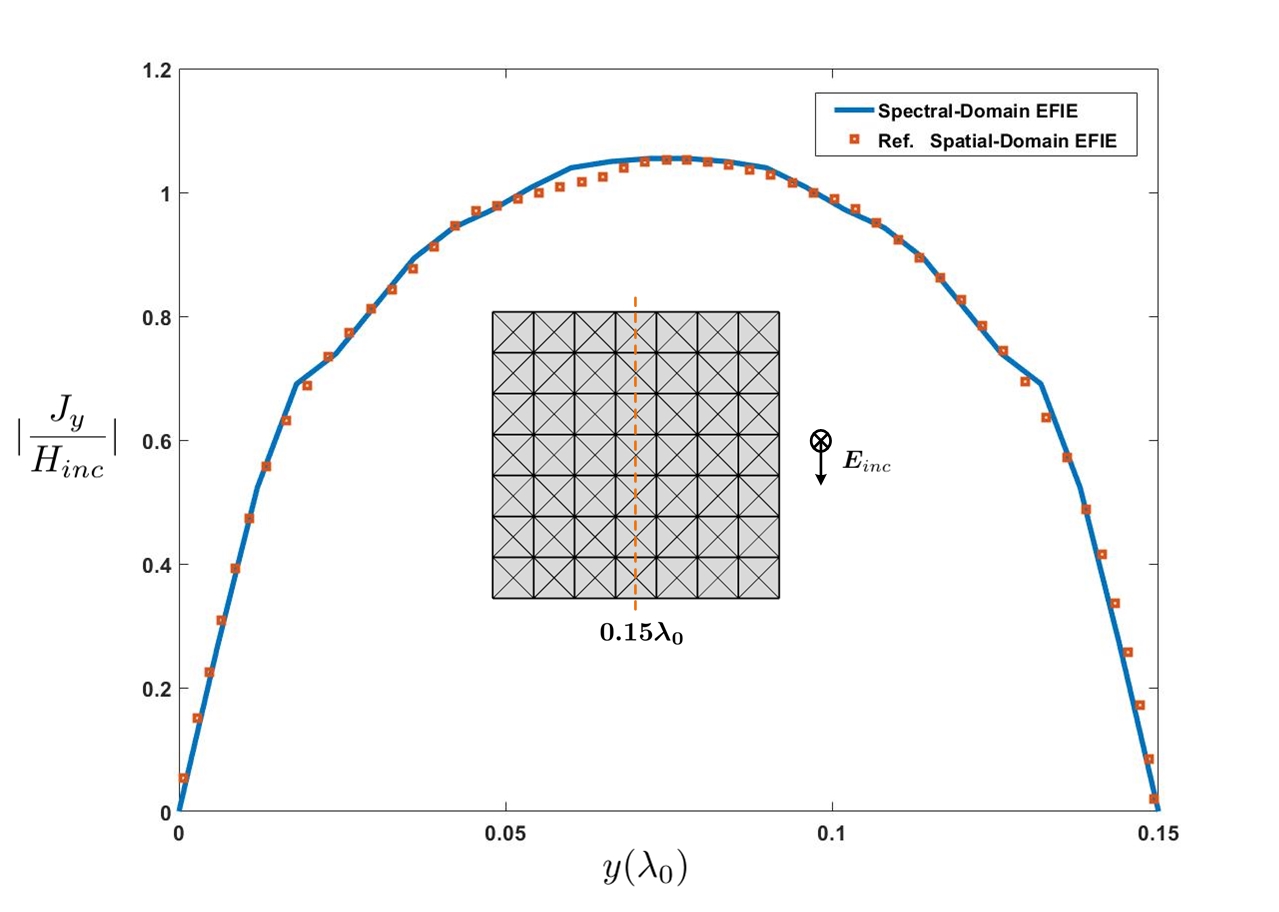}
	\caption*{ Fig. 1. Comparison of simulation results between the surface currrents obtained from the spatial-domain EFIE and spectral-domain EFIE of a PEC plate normally incident by a planar wave.}
	\label{fig:tl}
\end{figure}
 \begin{subequations}
\begin{eqnarray}
x_m = xk^m,\quad x_p = xk^p
\end{eqnarray}
\begin{eqnarray}
\alpha = \dfrac{2k^m e^{jx_p}\cos x_m}{jx_m},\quad
\beta = \dfrac{2k^m e^{jx_p}\sin x_m}{x_m}
\end{eqnarray}
\begin{equation}
	\begin{split}
&	\tilde{T}_{i+1}(x) = \\
&	\begin{cases}
	\alpha-\dfrac{2(i+1)\tilde{T}_i(x)}{jx_m}+\dfrac{i+1}{i-1}(\tilde{T}_{i-1}(x)-\alpha),\,& i = \textrm{even}\\
	\beta-\dfrac{2(i+1)\tilde{T}_i(x)}{jx_m}+\dfrac{i+1}{i-1}(\tilde{T}_{i-1}(x)-\beta),\, & i = \textrm{odd}
	\end{cases}
	\end{split}
\end{equation}
\begin{eqnarray}
	\tilde{T}_0(x) = \beta,\quad \tilde{T}_1(x) = \alpha - \dfrac{\beta}{jx_m}
\end{eqnarray}
\begin{eqnarray}
	\tilde{T}_2(x) = \dfrac{-4\alpha}{j x_m} + \beta(1-\frac{4}{x_m^2})\,.
\end{eqnarray}
\end{subequations}
\begin{figure}[t]
	\begin{minipage}{\columnwidth}
		\centering
		{\includegraphics[width=\columnwidth]{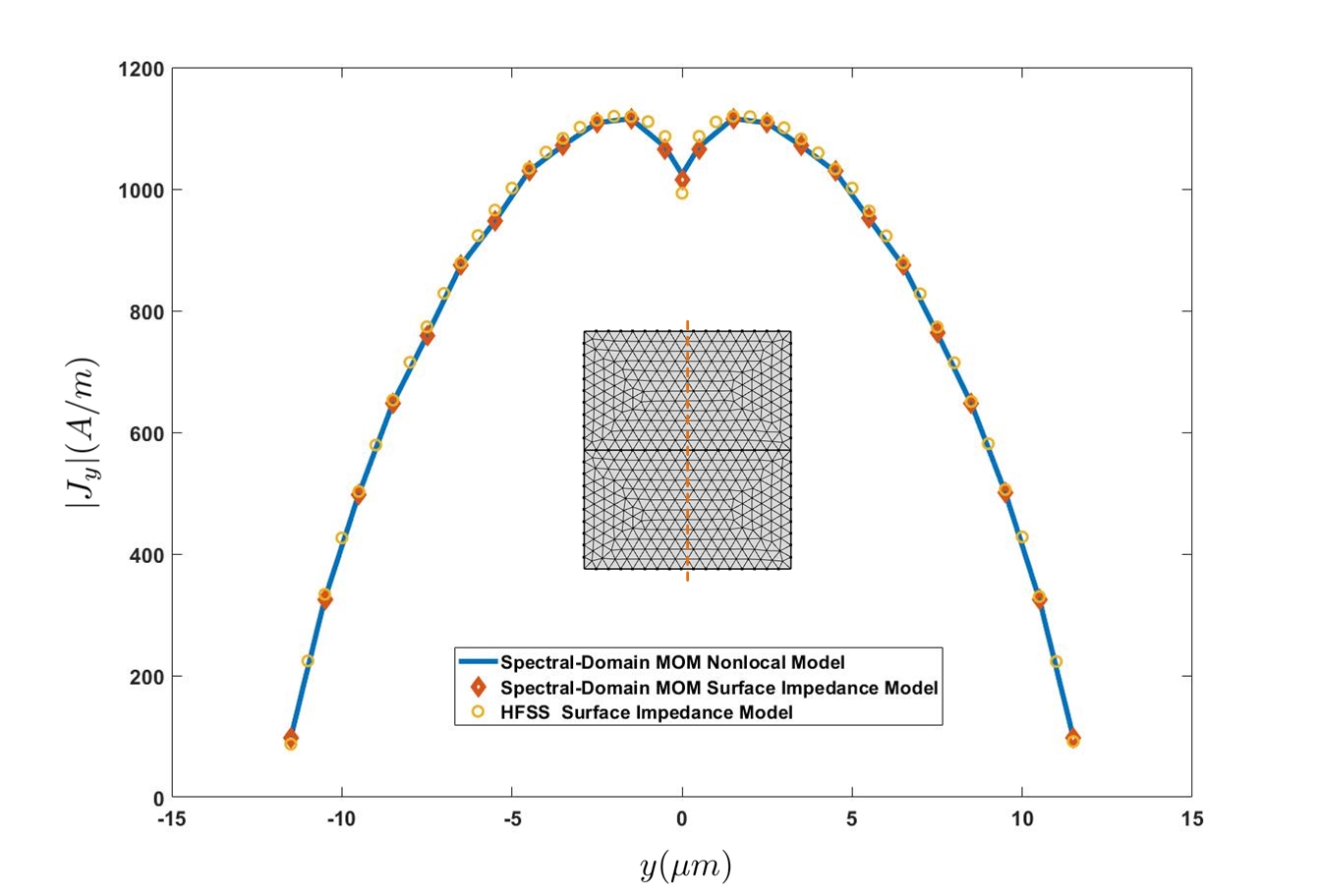}}
		\caption*{(a)}
	\end{minipage}
	\begin{minipage}{\columnwidth}
		\centering
		{\includegraphics[width=\columnwidth]{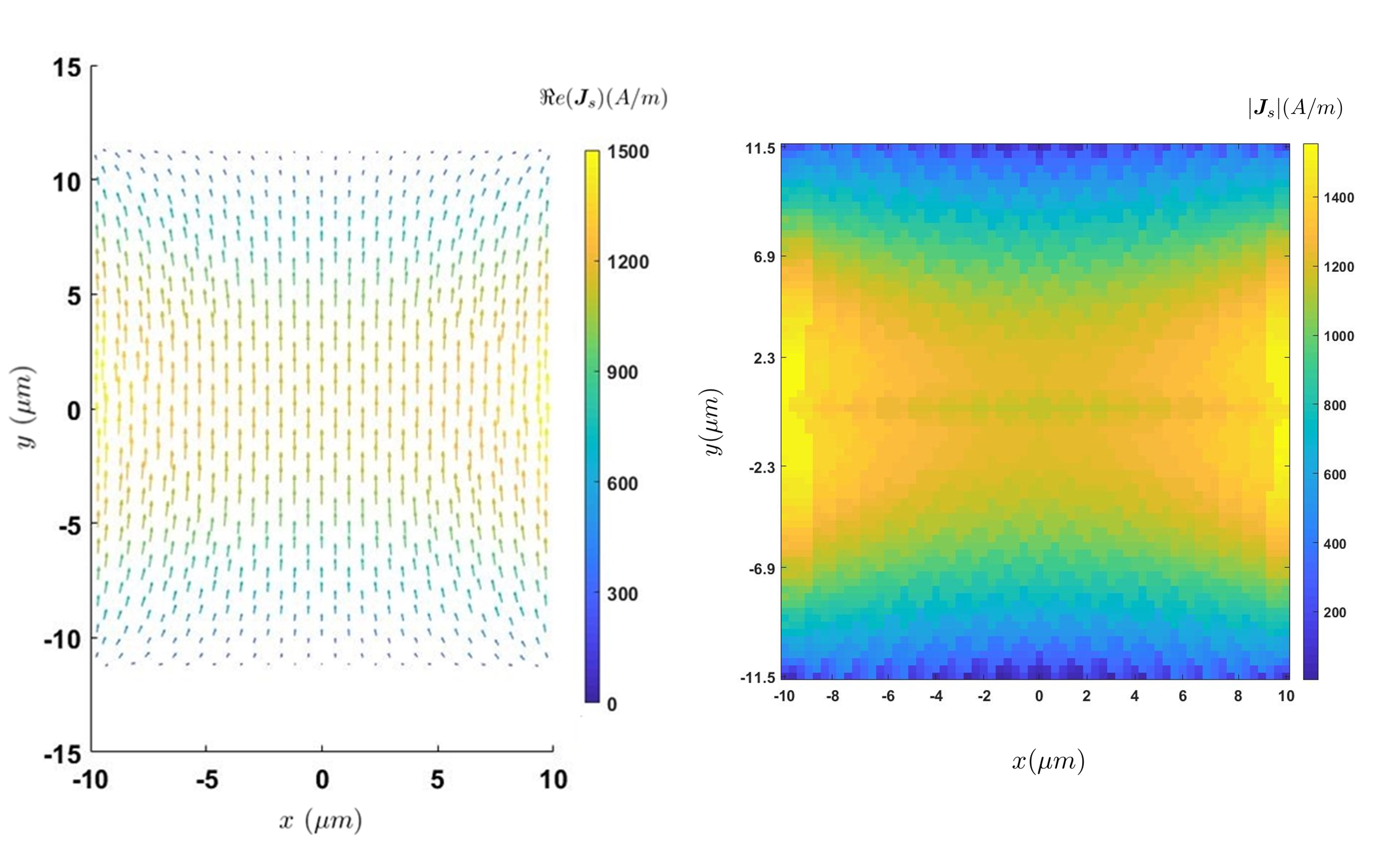}}
		\caption*{(b)}
	\end{minipage}
	\caption*{Fig. 2. (a) Scattered $|J_y|$ along the y-axis obtained by nonlocal and local graphene conductivity model respectively, using spectral-domain MOM and HFSS. (b) Snapshot of scattered surface currents $\Re e(\bm{J_s})$ at zero-phase timestamp, and magnitude of $|\bm{J}_s|$ obtained by nonlocal model.}
\end{figure}
The recursive relation starts from $i = 2,\,...,\,N_x-1$. If $x = 0$, alternately 
\begin{eqnarray}
	\tilde{T}_i(x) = 
	\begin{cases}
	\dfrac{2k^me^{jx_p}}{1-i^2},\,& i = \textrm{even}\\
	0,\,              & \text{otherwise}
	\end{cases}
\end{eqnarray}
	It is found the above recursive formulation is sensitive to the numerical roundoff errors for $x_m<5.0$ coresponding to small x or y offset of $r^\prime_n-r_m$. Therefore high numercial precison is required for the computational implementation. All the orders of $\tilde{T}_i(x)$ within the range of simulation space can be precompuated and saved in an interpolation table before the matrix filling.
\par
It is found an order of 32 Chebyshev polynomials is generally sufficient. Truncating bounds of $k_{x,y}$ are determined by the decadent rate of the spectral-domain RWG basis function, which is subjective to the edge length of the basis function. For all the simulations conducted in this work, we empirically choose
\begin{eqnarray}
	k_{max} = \dfrac{2\lambda_0 k_0}{l_{min}}
   \end{eqnarray}
where $l_{min}$ indicates the minimum edge length of the RWG basis function.
\section{Computational Results}
\subsection{Validation of Planar Wave Scattering on a PEC Plate}
To validate our code, our first example computes a planar wave normally incident on a flat PEC plate, which is previously considered in ~\cite{rao1982electromagnetic}. The PEC plate is placed in the free space and EFIE with free-space Green function in ~\cite{rao1982electromagnetic} is used as comparison. In Fig. 1 (a), the principle component of the surface currents is sampled along the y-axis shown as the orange dashed line. We find good agreement between our code and the result obtained from \cite{rao1982electromagnetic} in most of the data points. Although there seems to be a small discrepancy between the scattered field in the vicinity of the middle of the plate. This is likely due to the facts that the meshes used in \cite{rao1982electromagnetic} only consist of 60 triangles and are asymmetrically distributed, which result in discretization error.
\begin{figure*}[t]
	\centering
	\begin{minipage}{2\columnwidth}
		\centering
		\includegraphics[width=\columnwidth]{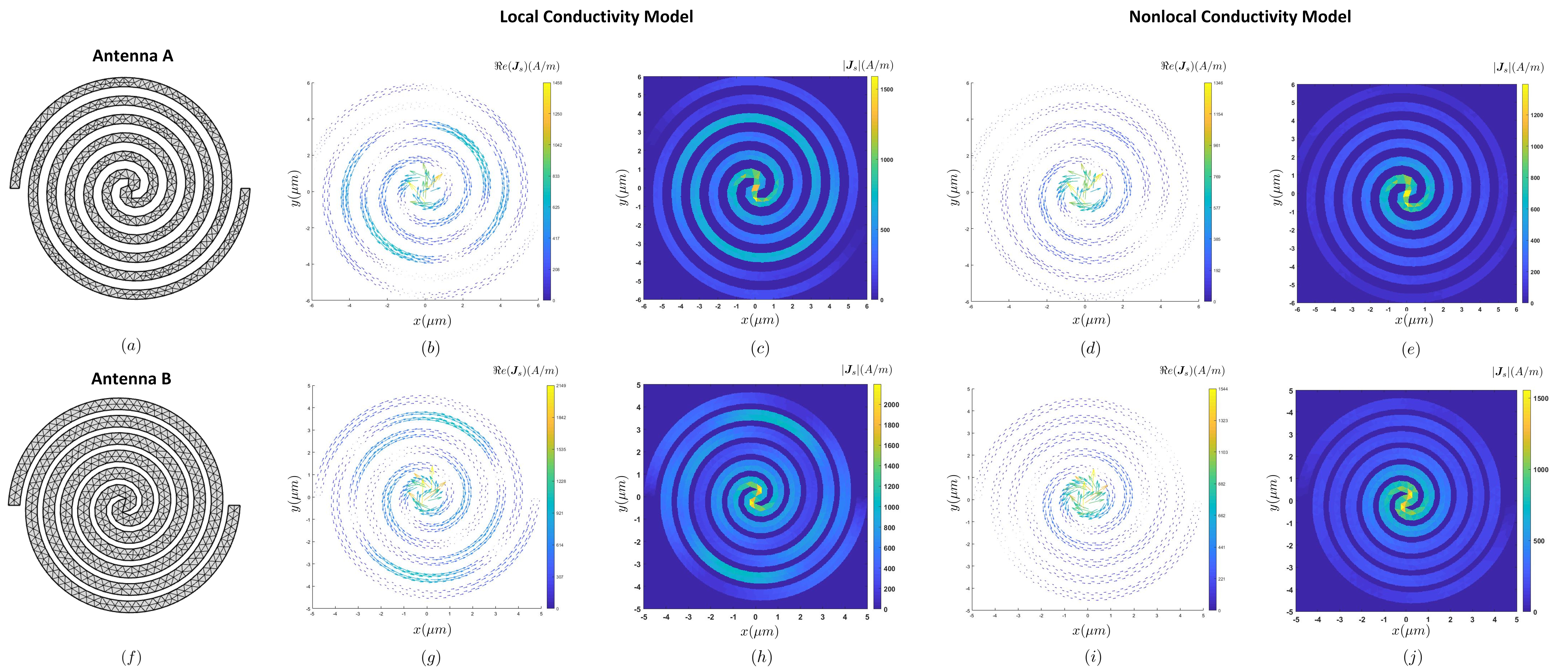}
	\end{minipage}
	\caption*{Fig. 3. (a) (f) The illustration and discretization meshes of the three-turn spiral antenna A, with width $0.5\mu m$, gap $0.5\mu m$, and antenna B, with width $0.5\mu m$, gap $0.5\mu m$, respectively. (b) (g) Scattered $\Re e(\bm{J}_s)(A/m)$ and 
		(c) (h) $|\bm{J}_s|(A/m)$ of antenna A and B respectively, using local surface impedance model. (d) (i) Scattered $\Re e(\bm{J}_s)(A/m)$ and (e) (j) $|\bm{J}_s|(A/m)$ of antenna A and B respectively, using nonlocal model.}
\end{figure*}

\subsection{Graphene Short Dipole Antenna}
In the second example, we compute a plasmonic electrically short antenna which is previously considered in \cite{dragoman2010terahertz}. The structure is consist of two graphene patches. 1V voltage is excited at the interface in the middle. The total length of the antenna is $23\ \mu m$, and the width is $20\ \mu \textrm{m}$. The lower space of the structure consists of an $\epsilon_r = 3.8$ substrate, and the upper space is free space. The graphene parameters are $\textrm{chemical potential}\ \mu_c = 0.2\, \textrm{eV},\, \textrm{relaxation time}\ \tau = 1\, \text{fs}$, $\textrm{temperature}=300\,\textrm{K},\ \textrm{and Fermi velocity}\ v_F = 10^6\ \textrm{m/s}$, and frequency of interest is 1 THz. The input impedance of the antenna obtained by the local and nonlocal conductivity model are $39.87+j\ 7.26\Omega$ and $39.69+j\ 6.37 \Omega$ respectively, which are almost identical. The current distribution of these two models along the y axis is also shown in Fig.~2 (a) and found to be identical. These results suggest that no notable nonlocal effect exists on this structure. This is due to the fact that the dipole antenna only excites standing waves, and the width of the patch is also large enough to ignore the quantum effect of graphene plasmons.
\begin{figure*}[t]
	\centering
	\begin{minipage}{2\columnwidth}
		\centering
		\includegraphics[width=\columnwidth]{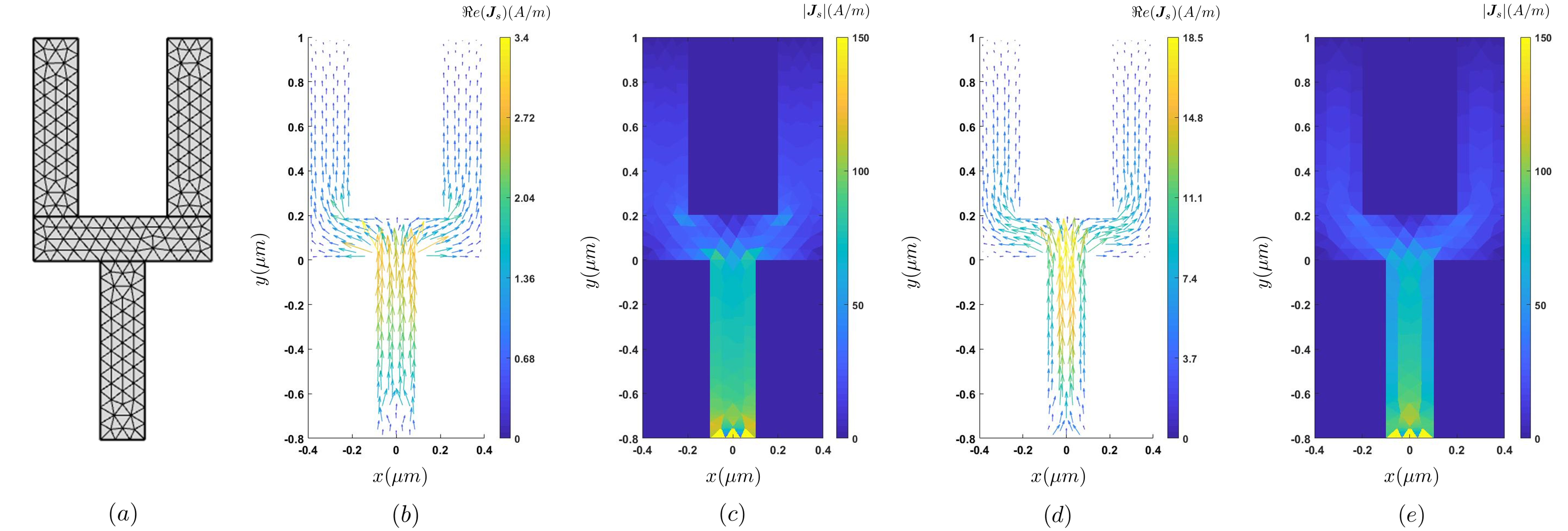}
	\end{minipage}
	\caption*{Fig. 4. (a) The illustration and discretization meshes of the two-way power divider. (b) Scattered $\Re e(\bm{J}_s)(A/m)$, and (c) $|\bm{J}_s|(A/m)$ using local surface impedance model. (d) Scattered $\Re e(\bm{J}_s)(A/m)$, and (e) $|\bm{J}_s|(A/m)$ using nonlocal model. }
	\label{figgra}
\end{figure*}
\subsection{Graphene Spiral Antennas}
In the third example, we consider an archimedean six-turn spiral graphene antenna place above a substrate of $\epsilon_r = 3.8$. The spiral geometry is illustrated in Fig.~3 (a) (f). We compute six different configurations of which the width and gap ranging from $0.25\ \mu m - 2\ \mu m$. The input impedance using both the local and nonlocal models are listed in TABLE 1. We observe when the width of the spiral antenna is narrower than $1\ \mu m$, the input impedance incorporating the nonlocal model obviously departs from the local model and reveals larger reactance values, while this phenomenon is less obvious for larger width of the stripline. In Fig.~3, the zero-phase snapshot and the magnitude of the surface currents for the case of the stripline width of $0.25\ \mu m$ are shown. The current distribution is different between these two models. It can be concluded that the width of the stripline plays an important role in the nonlocal effect. The shift of the antenna input impedance is attributed to the fact that the SPWs propagating on the spiral antenna is belonging to travelling wave with a wavelength much smaller than the free space. Since the surface conductivity of large wavenumber predicts a very different value from the lower wavenumber one, a signaficant deviation between these two models is expected for travelling-wave structures.

	\begin{table}
		\caption{\label{tab:t1}%
			Input Impedance of Graphene Spiral Antennas
		}
\begin{minipage}{\columnwidth}
			\centering
			\begin{tabular}{cccc}\hline\hline
				\\[-0.8em]
				\centering
				W ($\mu m$)&
				G ($\mu m$)&
				$Z_{in}\ (\Omega)$ \footnote{\ Input impedance of antennas using a local surface impedance model}&
				$Z_{in}\ ^*(\Omega)$ \footnote{\ Input impedance of antennas using a full-wavevector model}\\
				\\[-0.8em]
				\hline
				\hline
				\\[-0.5em]
				2 & 2& 917.18-j403.82& 899.11-j412.56\\
				2 & 1& 592.09-j354.79& 585.82-j355.40\\
				1 & 1& 1469.06-j348.08& 1443.83-j440.61\\
				1 & 0.5& 1242.19-j265.90& 1264.39-j372.97\\
				0.5 & 0.5& 1992.95-j476.71& 1832.35-j754.15\\
				0.5 & 0.25& 1432.69-j330.02& 1520.69-j774.95\\
				\\[-0.8em]
				\hline
			\end{tabular}
	\end{minipage}	
		\end{table}

\subsection{Graphene Power Divider}
In the third example, we consider a two-way power divider made of graphene stripline which is also previously considered in \cite{li2018discontinuous}. The width of the graphene is $0.2\ \mu m$, and the length of each branch is $0.8\ \mu m$. The geometry of the device is illustrated in Fig.~4 (a). The device is placed above a substrate of $\epsilon_r = 3.8$.  Since the stripline is very narrow, strong nonlocal effect is expected to be seen. We observe from Fig.~4 (c) (e) that the current distribution of the nonlocal model is concentrated on the center of the stripline while the local model predicts uniformed distributed surface currents. The results indicate that these two models of conductivity predict a different dominated propagation mode of the graphene stripline \cite{nikitin2011edge}, and confirms the necessity of incorporating the nonlocal model in miniaturized-size graphene devices.

\section{Conclusion}
Formulation and computational methods of the spectral-domain method of moments using the Rao-Wilton-Glisson subdomain basis function are proposed to model anisotropic and spatially dispersive graphene patch embedded in planarly layered uniaxial media. Several computational examples that rigorously model the surface plasmonic waves excited on graphene patches exhibiting significant spatial dispersion are demonstrated.
 
\appendices
\section{Graphene Surface Conductivity Tensor}
\label{a3g}
The surface conductivity tensor used in this paper is derived from the semiclassical Boltzmann transport equation under both the relaxation-time approximation and the Bhatnagar-Gross-Krook model, which models the intraband transitions of graphene and includes the spatial dispersion for transverse wavevector. The closed-form  expressions below are strictly correct only when $\mu_c = 0\, \text{eV}$. However, numerical results confirm that simulation from the closed-form expressions show a very good agreement with the exact numerical integration over the first Brillouin zone using the tight-binding electron dispersion relation \cite{lovat2013semiclassical}.
\begin{subequations}
	\begin{eqnarray}
	\sigma_{xx}^{BGK} (k_x,k_y) = \gamma \dfrac{I_{\phi_{xx}} + \gamma_D \Delta k_y (I_{\phi_{xx}}k_y - I_{\phi_{yx}}k_x)}{D_{\sigma}}\\
	\sigma_{xy}^{BGK} (k_x,k_y) = \gamma \dfrac{I_{\phi_{xy}} + \gamma_D \Delta k_y (I_{\phi_{xy}}k_y - I_{\phi_{yy}}k_x)}{D_{\sigma}}\\
	\sigma_{yx}^{BGK} (k_x,k_y) = \gamma \dfrac{I_{\phi_{yx}} + \gamma_D \Delta k_x (I_{\phi_{yx}}k_x - I_{\phi_{xx}}k_y)}{D_{\sigma}}\\
	\sigma_{yy}^{BGK} (k_x,k_y) = \gamma \dfrac{I_{\phi_{yy}} + \gamma_D \Delta k_x (I_{\phi_{yy}}k_x - I_{\phi_{xy}}k_y)}{D_{\sigma}}
	\end{eqnarray}
\end{subequations}
with
\begin{subequations}
	\begin{eqnarray}
	&I_{\phi_{xx}}(\omega,k_x,k_y) = \notag \\ &2\pi\dfrac{v_f^2 k_y^2 k_t^2 R - \alpha v_f k_x k_q^2 -\alpha^2 k_q^2(1-R)}{v_f^2(\alpha+v_fk_x)k_t^4}\\
	&I_{\phi_{xy}}(\omega,k_x,k_y) = I_{\phi_{yx}}(k_x,k_y) = \notag \\ &-2\pi k_x k_y\dfrac{v_f^2 k_t^2 R + 2\alpha v_f k_x +2\alpha^2 (1-R)}{v_f^2(\alpha+v_fk_x)k_t^4}\\
	&I_{\phi_{yy}}(\omega,k_x,k_y) = \notag\\& 2\pi\dfrac{v_f^2 k_x^2 k_t^2 R + \alpha v_f k_x k_q^2 +\alpha^2 k_q^2(1-R)}{v_f^2(\alpha+v_fk_x)k_t^4}
	\end{eqnarray}
\end{subequations}
and
\begin{subequations}
	\begin{eqnarray}
&	\gamma = -j \frac{e^2 k_B T}{\pi^2 \hbar^2} \log\{2[1+\cosh(\frac{\mu_c}{k_B T})] \}	\\
&\gamma_D = j \frac{v_f}{2\pi \omega \tau},\quad	D_{\sigma} = 1+ \gamma_D \Delta k_t^2\\
&	\Delta = \frac{-2\pi}{v_f k_t^2}(1-\frac{\alpha}{\sqrt{\alpha^2 - v_f^2 k_t^2}})\\
&	R(k_x,k_y) = \dfrac{\alpha + v_f k_x}{\sqrt{\alpha^2 - v_f^2 k_t^2}},\quad \alpha = \omega - \frac{j}{\tau}\\
&	k_t = \sqrt{k_x^2 + k_y^2},\quad k_q = \sqrt{k_x^2 - k_y^2}
	\end{eqnarray}
\end{subequations}
where $\omega$ is the angular frequency, $k_B$ is Boltzmann constant, T is temperature, $e$ is electron charge, $\hbar$ is reduced Planck constant, $\tau$ the phenomenological relaxation time, $v_f$ the
Fermi velocity, and $\mu_c$ is graphene’s chemical potential. 




\ifCLASSOPTIONcaptionsoff
\newpage
\fi
\bibliographystyle{IEEEtran}
%
\bibliography{btxdoc}

%





\end{document}